\documentclass[aps,pra,showpacs,groupedaddress,superscriptaddress,twocolumn]{revtex4}
\usepackage{graphicx}
\usepackage{latexsym}
\usepackage{makeidx}
\usepackage{amsmath}
\usepackage{amssymb}
\usepackage{graphicx}

\input{tcilatex}

\begin{document}

\title{Photon generation from vacuum in non-stationary circuit QED}

\begin{abstract}
We study theoretically the non-stationary circuit QED system in which the
artificial atom transition frequency has a small periodic modulation in
time, prescribed externally. We show that, in the dispersive regime, when
the modulation periodicity meets the `resonances', the dynamics may be
described by the dynamical Casimir effect, Jaynes-Cummings or
Anti-Jaynes-Cummings effective Hamiltonians. In the resonant atom-cavity
regime, under the modulation `resonance' the dynamics resembles the behavior
of the dynamical Casimir effect in a vibrating cavity containing a resonant
two-level atom, and entangled states with two photons can be created from
vacuum. Thus, an analog of the dynamical Casimir effect may be simulated in
circuit QED, and several photons, as well as entangled states, can be
generated from vacuum due to the anti-rotating term in the Rabi Hamiltonian.
\end{abstract}

\author{A. V. Dodonov}
\affiliation{Departamento de F\'{\i}sica, Universidade Federal de S\~{a}o Carlos, S\~{a}o
Carlos, 13565-905, S\~{a}o Paulo, Brazil}
\author{L. C. C\'{e}leri}
\affiliation{Universidade Federal do ABC, Centro de Ci\^{e}ncias Naturais e Humanas, R.
Santa Ad\'{e}lia 166, Santo Andr\'{e}, 09210-170, S\~{a}o Paulo, Brasil.}
\author{F. Pascoal}
\affiliation{Departamento de F\'{\i}sica, Universidade Federal de S\~{a}o Carlos, S\~{a}o
Carlos, 13565-905, S\~{a}o Paulo, Brazil}
\author{M. D. Lukin}
\affiliation{Physics Department, Harvard University, Oxford St., Cambridge, MA 02138, USA}
\affiliation{ITAMP, Harvard-Smithsonian Center for Astrophysics, Cambridge, MA 02138, USA}
\author{S. F. Yelin}
\affiliation{ITAMP, Harvard-Smithsonian Center for Astrophysics, Cambridge, MA 02138, USA}
\affiliation{Department of Physics, University of Connecticut, Storrs, CT 06269, USA}
\maketitle

\section{Introduction}

Over the last few decades nonstationary processes in cavities have received
considerable attention. One such process is the nonstationary or dynamical
Casimir Effect (DCE) -- in particular, the creation of photons from vacuum,
or another initial field state, in cavity whose geometry \cite%
{PRA50-1027,PRA50-1027a,PRA50-1027b,Law,Law1} or material properties \cite%
{Law,Law1,jj,jj1,jj2,jj3} have a periodic time dependence, with modulation
frequency equal to twice the unperturbed field eigenfrequency. Nowadays, DCE
in cavities is a well studied problem, with a variety of theoretical
predictions concerning the number and the statistics of created photons, as
well as the influence of detuning, dissipation, boundary conditions,
geometry and non-periodicity of the modulation (see \cite%
{JPB-S,book,JRLR26-445} for an extensive list of references). To date, DCE
has not been observed in laboratory, however several concrete proposals have
appeared over the last years \cite%
{JPA39-6271,JPA39-6271a,onofrio,JPA39-6271b,d2,PRB72-115303,mme}, with some
of them being currently implemented experimentally \cite{EPL70-754}.

The interest in nonstationary processes in cavities reappeared over the last
5 years due to the progress in the field of Cavity Quantum Electrodynamics
(cavity QED \cite{S298-1372}) in the condensed matter systems, e.g.
semiconductor quantum dots \cite{N432-197,N432-200}, polar molecules \cite%
{NP2-636} and superconducting circuits \cite{N431-159,N431-162,PRL96-127006}
coupled to resonators, the latest architecture known as \textit{circuit} QED
\cite{N431-162,revvv}. In cavity QED, the effective two-level atom is
coupled to the field inside the resonator via the dipole interaction,
allowing for observation of the light-matter interaction at the level of
single photons and single atoms. A new ingredient in the solid state cavity
QED is the possibility of engineering and manipulating the properties of the
artificial atom and the resonator \cite{N431-162,xx5,xx6}, as well as the
interaction strength between them, either during the fabrication or \textit{%
in situ}.

Recently, the strong resonant and the strong dispersive coupling limits
between the artificial atom and a single cavity mode were observed
experimentally in circuit QED \cite{N445-515} and other solid state
architectures \cite{N432-197,N432-200}. Moreover, the single photon source
\cite{N449-7160}, single artificial-atom maser \cite{xx4}, multiphoton Fock
states \cite{medv} and interaction between two artificial atoms (qubits)
\cite{xx6,N449-443} were implemented experimentally, among many other
important achievements. Besides, the circuit QED architecture benefits from
robust read-out schemes of the atomic internal state and the resonator field
state \cite{medv,PRA69-062320,r1,r2,PRA75-032329}, relatively low
dissipative losses \cite{N445-515}, state preparation techniques \cite{gp}
and real time manipulation of the atomic transition frequency via electric
and magnetic fields \cite{N431-162,xx6,medv} or non-resonant microwave
fields \cite{N449-443,PRA75-032329}.

Harnessing the tunability of the atomic transition frequency, the
realization of the Landau-Zener sweeps, with the atomic frequency increasing
linearly in time, was proposed in circuit QED \cite{lz}, allowing for
generation of single photons and entangled states. In a similar direction,
in \cite{d2,PRB72-115303,d1} the implementation of DCE in semiconductor
cavity QED was considered, using the periodic time-dependence of the
atom-cavity coupling parameter (the vacuum Rabi frequency). It was shown
that there is a substantial photon production from vacuum when the vacuum
Rabi frequency is modulated in time with certain `resonant' frequencies \cite%
{d2}. A preliminary theoretical study of the feasibility of realizing the
DCE with a quantum flux qubit in superconducting quantum nanocircuits, as
well as the detection of the generated photons, was reported in \cite{sss}.
On the other hand, the possibility of controlling the atomic frequency and
detecting the atomic internal state is currently being used to
couple/decouple one or several qubits to/from the cavity mode in order to
implement quantum logic operations \cite{xx6,N449-443,PRA75-032329}.

Here we study the nonstationary cavity QED architecture, in which a single
cavity mode is coupled to a single artificial atom whose transition
frequency has a small periodic modulation in time prescribed externally.
Such a control over the transition frequency, with compatible modulation
periodicity, may be achieved in circuit QED with present or near-future
technology \cite{xx6,medv,PRA75-032329}. We show that, in the dispersive
regime, under the `resonance' conditions one obtains completely different
effective regimes for the system dynamics, which may be approximately
described by the Anti-Jaynes-Cummings (AJC), Jaynes-Cummings (JC) or the
dynamical Casimir effect (DCE) Hamiltonians with adjustable parameters.
Moreover, in the resonant atom-cavity regime, the dynamics resembles the
behavior of the DCE in a cavity with oscillating boundaries containing a
resonant two-level atom (detector) \cite{pp}, and entangled states with up
to two photons can be generated from vacuum.

Thus, we demonstrate the possibility of simulating the DCE in circuit QED
using a single non-stationary atom, instead of a macroscopic dielectric
medium as in \cite{jj}. As applications, it may be possible to create
excitations, either photonic or atomic, from the initial vacuum state $%
|g,0\rangle $, generate nonclassical states of light and realize transitions
between the states $\left\{ |g,m\rangle ,|e,m\pm 1\rangle \right\} $ in the
dispersive regime. Here $|g\rangle $ and $|e\rangle $ stand for the atomic
ground and excited states, respectively, and $|m\rangle $ for the Fock state
of the cavity field. A related problem was recently studied in \cite{today},
where it was suggested that lasing behavior and the creation of a highly
non-thermal population of the oscillator, as well as the cooling, could be
implemented using an analogous scheme in the near future.

\section{Nonstationary circuit QED}

We assume that the atomic transition frequency $\Omega \left( t\right) $ may
be described as the sum of two terms. The first term $\Omega _{0}$ describes
the bare atomic frequency and the second term represents a small modulation
amplitude $\varepsilon \ll \Omega _{0}$ multiplied by a periodic function of
time $f_{t}$ prescribed externally%
\begin{equation}
\Omega \left( t\right) =\Omega _{0}+\varepsilon f_{t},\quad
f_{t}=\sum_{k=0}^{\infty }\left( s_{k}\sin k\eta t+c_{k}\cos k\eta t\right) .
\label{ft}
\end{equation}%
Here $\eta $ is the modulation frequency and $\left\{ s_{k},c_{k}\right\} $
form a set of coefficients describing an arbitrary periodic time-dependence
of $f_{t}$. We suppose that the cavity frequency $\omega $ and the
atom-cavity coupling parameter $g_{0}$ are constant, so at the `sweet spot'
in solid state cavity QED the system is described by the Rabi Hamiltonian
(RH) \cite{PRA69-062320}
\begin{equation}
H=H_{0}\left( t\right) +g_{0}\left( a+a^{\dagger }\right) \left( \sigma
_{+}+\sigma _{-}\right) ,  \label{Rabi}
\end{equation}%
where $a$ ($a^{\dagger }$) is the cavity annihilation (creation) operator, $%
\sigma _{+}=|e\rangle \langle g|$ and $\sigma _{-}=|g\rangle \langle e|$.
The free Hamiltonian is%
\begin{equation*}
H_{0}\left( t\right) =\omega n+\frac{\Omega \left( t\right) }{2}\sigma _{z},
\end{equation*}%
where $n=a^{\dagger }a$, $\sigma _{z}=|e\rangle \langle e|-|g\rangle \langle
g|$ and we assume $\hbar =1$. In the stationary case, $\varepsilon =0$, one
may perform the Rotating Wave Approximation (RWA) and obtain the standard
Jaynes-Cummings (JC) Hamiltonian \cite{Klimov}, which has been verified in
several experiments over the last few years \cite%
{N431-162,N445-515,climbing,rempe,ss}. However, in the non-stationary case,
as well as under strong dephasing noise \cite{TW}, the anti-rotating term $%
\left( a\sigma _{-}+a^{\dagger }\sigma _{+}\right) $ cannot always be
eliminated. Moreover, it is responsible for producing an analog of the DCE
and creating photonic and atomic excitations from vacuum under modulation
`resonance' conditions, as shown below.

In the interaction picture with respect to $H_{0}\left( t\right) $ the
interaction Hamiltonian reads
\begin{equation}
H_{I}=g_{0}\left( e^{i\Xi _{-}}a\sigma _{+}+e^{i\Xi _{+}}a^{\dagger }\sigma
_{+}+h.c.\right) ,  \label{a1}
\end{equation}%
where h.c. stands for the Hermitian conjugate and $\Xi _{\pm }\equiv
\int_{0}^{t}d\tau \left[ \Omega \left( \tau \right) \pm \omega \right] .$
All the information about the dynamics of the system is contained in the
time-dependent coefficients $\exp (i\Xi _{\pm })$, which may be
significantly simplified by adjusting the modulation frequency $\eta $ in
order to achieve the `resonances'. We have explicitly
\begin{equation}
g_{0}e^{i\Xi _{\pm }}=ge^{i\Delta _{\pm }t}\sum_{l=0}^{\infty }\frac{1}{l!}%
\left[ \frac{\varepsilon }{\eta }\sum_{k=1}^{\infty }\left( \Lambda
_{k}e^{-ik\eta t}-\Lambda _{k}^{\ast }e^{ik\eta t}\right) \right] ^{l},
\label{eee}
\end{equation}%
where we defined a complex coupling constant $g\equiv g_{0}\exp \left[
i\left( \varepsilon /\eta \right) \sum_{k=1}^{\infty }k^{-1}s_{k}\right] $
and parameters%
\begin{equation*}
\Lambda _{k}\equiv -\frac{c_{k}+is_{k}}{2k},\quad \Delta _{\pm }\equiv
\Omega _{0}+\varepsilon c_{0}\pm \omega .
\end{equation*}

\section{AJC and JC resonances}

The `Anti-Jaynes-Cummings' (AJC) resonance occurs for%
\begin{equation*}
\eta =\eta _{AJC}\equiv \Delta _{+}-\xi ,
\end{equation*}%
where $\left\vert \xi \right\vert \ll \eta $ is a small `resonance shift'.
Assuming a reasonable experimental condition $\varepsilon /\eta \ll 1$ we
expand (\ref{eee}) to the first order in $\varepsilon /\eta $ and make the
RWA in (\ref{a1}), obtaining
\begin{equation}
H_{I}\simeq g\left( \theta e^{i\xi t}a^{\dagger }\sigma _{+}+e^{i\Delta
_{-}t}a\sigma _{+}\right) +h.c.,  \label{resul}
\end{equation}%
where the AJC dimensionless coupling is
\begin{equation}
\theta \equiv \Lambda _{1}\varepsilon /\eta \,.  \label{theta}
\end{equation}

In the resonant regime, $\left\vert \Delta _{-}\right\vert /g_{0}\ll 1$, we
apply the method of slowly varying amplitudes to the Hamiltonian (\ref{resul}%
), repeating the procedure employed originally for studying the photon
generation from vacuum due to the DCE in a vibrating cavity containing a
stationary two-level atom \cite{pp}. We find that for the initial state $%
|g,0\rangle $ the photon generation occurs for two values of the resonance
shift%
\begin{equation*}
\xi =\xi _{\pm }\equiv \Delta _{-}/2\pm \sqrt{2}g_{0}
\end{equation*}%
and one gets the following non-zero probabilities $P_{x,m}$, with $x=\left\{
e,g\right\} $ denoting the atomic state and $m$ the photon number%
\begin{eqnarray}
P_{g,0} &\approx &\cos ^{2}\left( \chi t\right) ,\quad P_{e,1}\approx \sin
^{2}\left( y+q\right) \sin ^{2}\left( \chi t\right) ,\quad   \notag \\
P_{g,2} &\approx &\cos ^{2}\left( y-q\right) \sin ^{2}\left( \chi t\right) .
\label{PPP}
\end{eqnarray}%
Here%
\begin{equation*}
\chi \approx g_{0}\left\vert \theta \right\vert \sin \left( y+q\right)
,\quad \tan y\approx \left[ \frac{2\sqrt{2}g_{0}+\Delta _{-}}{2\sqrt{2}%
g_{0}-\Delta _{-}}\right] ^{1/2}
\end{equation*}%
and $q=0$ ($\pi /2$) for $\xi _{-}$ ($\xi _{+}$). Thus, in the resonant
regime, when the atomic transition frequency is modulated with the frequency
$\Delta _{+}-\xi _{\pm },$ a superposition of states $|g,0\rangle $, $%
|e,1\rangle $ and $|g,2\rangle $ is created from the initial vacuum state $%
|g,0\rangle $, and the probability of exciting the atom is limited by $\sin
^{2}y\approx 1/2$.

We illustrate this behavior in Fig. 1a, where we show the exact dynamics of $%
P_{g,0}$, $P_{e,1}$ and $P_{g,2}$ \textit{vs}. time ($t$) for the AJC
resonance, $f_{t}=\sin \eta _{AJC}t$ with $\xi =\xi _{-}$, using the
parameters $g_{0}/\omega =4\cdot 10^{-2}$, $\Delta _{-}=g_{0}/10$ and $%
\varepsilon /\omega =10^{-1}$. \ This dynamics resembles the one occurring
in the context of DCE \cite{pp}, where a resonant (stationary) two-level
atom or detector is fixed inside a cavity whose boundary is oscillating with
the frequency close to $2\omega $. In both cases not more than two photons
can be created from the vacuum state $|g,0\rangle $ and the probability of
exciting the atom is limited by the value $1/2$. This similarity is not
surprising, since in this case the modulation frequency $\Delta _{+}\approx
2\omega $, and the atom plays the role of the two-level detector and the
cavity modulating mechanism (via the atom-cavity coupling) at the same time.
\begin{figure}[tbp]
\begin{center}
\includegraphics[width=.49\textwidth]{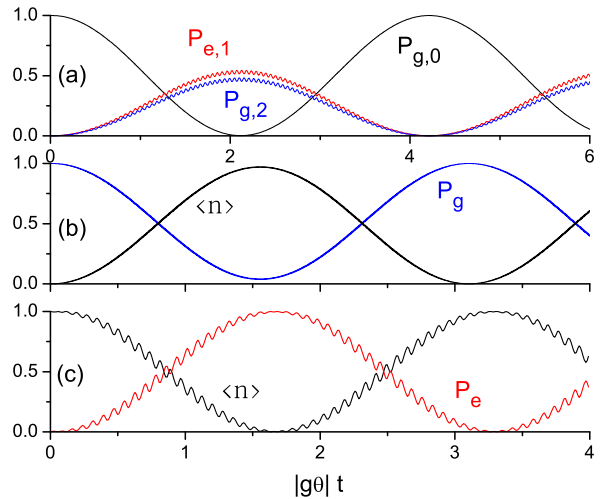} {}
\end{center}
\caption{\textbf{a)} Exact dynamics of $P_{g,0}$, $P_{e,1}$ and $P_{g,2}$
\textit{vs.} time in the resonant atom-cavity regime under the AJC
resonance. \textbf{b)} Exact dynamics of $\left\langle n\right\rangle $, $%
P_{g}$ and $P_{e}$ in the dispersive regime for the AJC resonance and
\textbf{c)} JC resonance.}
\end{figure}

In the dispersive regime, $g_{0}\sqrt{\left\langle n\right\rangle }%
/\left\vert \Delta _{-}\right\vert \ll 1$, where $\left\langle
n\right\rangle $ is the mean photon number, the Hamiltonian (\ref{resul})
may be approximated by \cite{Klimov}%
\begin{equation*}
H_{I}^{\left( 1\right) }\simeq \left( g\theta e^{i\xi t}a^{\dagger }\sigma
_{+}+h.c.\right) +\delta \left( n+1/2\right) \sigma _{z}~,
\end{equation*}%
where
\begin{equation*}
\delta =g_{0}^{2}/\Delta _{-}
\end{equation*}%
is the dispersive shift. In a rotating frame we obtain the effective AJC
Hamiltonian%
\begin{equation}
H_{AJC}\simeq \left[ \xi +\delta \left( 1+2n\right) \right] \frac{\sigma _{z}%
}{2}+\left( g\theta a^{\dagger }\sigma +h.c.\right)   \label{AJC}
\end{equation}%
By adjusting the resonance shift $\xi $ \cite{foot} in order to make $%
\left\vert \xi +\delta \left( 1+2n\right) \right\vert \ $small compared to $%
\left\vert g\theta \right\vert $, one obtains the resonant AJC Hamiltonian.
From the physical point of view, the external modulation supplies the energy
$\omega +\Omega _{0}$ necessary to create one photon and one atomic
excitation simultaneously. Thus, one can create the superposition of states $%
|e,1\rangle $ and $|g,0\rangle $ starting from the initial state $%
|g,0\rangle $, as illustrated in Fig. 1b, where we show the exact dynamics
of $\left\langle n\right\rangle $ and $P_{g}$ for the AJC resonance, $%
f_{t}=\sin \eta _{AJC}t$ with $\xi =-2\delta $, using the experimental
circuit QED parameters $\Omega _{0}/\omega =1.4$, $g_{0}/\omega =2\cdot
10^{-2}$ and assuming $\varepsilon /\omega =0.2$.

Also in the dispersive regime, the `Jaynes-Cummings' (JC) resonance occurs
for%
\begin{equation*}
\eta =\eta _{JC}\equiv \left\vert \Delta _{-}\right\vert -\xi .
\end{equation*}%
For positive $\Delta _{-}$ we get the effective JC Hamiltonian%
\begin{equation}
H_{JC}\simeq \left[ \xi +\delta \left( 1+2n\right) \right] \frac{\sigma _{z}%
}{2}+\left( g\theta \,\,a\sigma _{+}+h.c.\right) .  \label{JC}
\end{equation}%
If $\Delta _{-}$ is negative, we obtain the same effective Hamiltonian upon
replacements $\theta \rightarrow -\theta ^{\ast }$ and $\xi \rightarrow -\xi
$ in Eq. (\ref{JC}). Thus, by employing the JC resonance and adjusting the
resonance shift $\xi $, one may couple the subspaces $\left\{ |g,m\rangle
,|e,m-1\rangle \right\} $ when the atom and the field are far off-resonant,
since the external modulation supplies the energy difference $\left\vert
\omega -\Omega _{0}\right\vert $ necessary to couple the atom and the cavity
field. This behavior is illustrated in Fig. 1c, where we show the exact
dynamics of $\left\langle n\right\rangle $ and $P_{e}$ for the JC resonance,
$f_{t}=\sin \eta _{JC}t$ with $\xi =-2\delta $, for the initial state $%
|g,1\rangle $ and the parameters of Fig. 1b. Moreover, one could engineer
entangled states with several photons from the initial vacuum state $%
|g,0\rangle $ by alternating between the AJC and JC resonances and
controlling the time interval and the resonance shift of each resonance.

\section{DCE resonance}

In the dispersive regime, the `dynamical Casimir effect' (DCE) resonance
occurs for%
\begin{equation*}
\eta =\eta _{DCE}\equiv 2\omega -2\xi .
\end{equation*}%
Performing the RWA in the interaction Hamiltonian (\ref{a1}), in a rotating
frame we obtain the time-independent Hamiltonian consisting of the JC
Hamiltonian plus the AJC term multiplied by the adjustable coupling $g\theta
$%
\begin{equation}
H_{I}^{\left( 1\right) }\simeq \xi n+\frac{\Delta _{-}+\xi }{2}\sigma
_{z}+\left( ga\sigma _{+}+g\theta a^{\dagger }\sigma _{+}+h.c.\right) ~.
\label{vvv}
\end{equation}%
We may obtain an effective Hamiltonian by applying a sequence of small
unitary transformations \cite{Klimov} on (\ref{vvv}) and performing the
Hausdorff expansion after each step. Assuming that $\theta \sim O\left(
\left\vert g/\Delta _{-}\right\vert \right) $ we apply the `rotating'
unitary transformation
\begin{equation*}
U_{r}=\exp \left[ \left( ga\sigma _{+}-h.c.\right) /\Delta _{-}\right]
\end{equation*}%
followed by the `anti-rotating' one
\begin{equation*}
U_{a}=\,\exp \left[ \left( g\theta a^{\dagger }\sigma _{+}-h.c.\right)
/\left( \Delta _{-}+2\xi \right) \right] \,
\end{equation*}%
to obtain the effective Hamiltonian $H_{eff}=U_{a}U_{r}H_{I}^{\left(
1\right) }U_{r}^{\dagger }U_{a}^{\dagger }$, which in a rotating frame reads%
\begin{eqnarray}
H_{eff} &\simeq &\left( \xi +\delta \sigma _{z}\right) n+\delta \sigma
_{z}\left( \theta ^{\ast }a^{2}+h.c.\right)   \label{PDC} \\
&-&\frac{2\delta }{\Delta _{-}}\left( ge^{i\Delta _{-}t}an\sigma
_{+}+h.c.\right) +O(\left\vert g/\Delta _{-}\right\vert ^{3}).  \notag
\end{eqnarray}

The first two terms of the effective Hamiltonian (\ref{PDC}) form the DCE
part and the remaining terms represent the corrections, whose leading term
(oscillating with high frequency $\sim \left\vert \Delta _{-}\right\vert $)
describes the nonresonant photon absorption by the atom. These corrections
become relevant when the third term becomes large, so for initial times
(roughly for $g_{0}\sqrt{\left\langle n\right\rangle }/\left\vert \Delta
_{-}\right\vert \ll 1$) their contribution is relatively small and $\sigma
_{z}$ becomes approximately a constant. If the atom is initially in the
ground state, the Eq. (\ref{PDC}) becomes the DCE Hamiltonian%
\begin{equation}
H_{DCE}\simeq \left( \xi -\delta \right) n-\delta \left( \theta ^{\ast
}a^{2}+h.c.\right) .  \label{DCE}
\end{equation}%
For the atom initially in the excited state, a similar effective Hamiltonian
is obtained under substitution $\delta \rightarrow -\delta $. Therefore, by
adjusting the frequency shift to $\xi =\pm \delta $, depending on the
initial atomic state \footnote{%
If initially the atom is in the superpositions of states $|g\rangle $ and $%
|e\rangle $, we may choose any of the signs to obtain photon generation.
However, the photon generation is optimized if initially the atom is exactly
in $|g\rangle $ or $|e\rangle $.}, we have photon pairs creation from vacuum
and field amplification due to an analog of the DCE.

Here the DCE is simulated by the atomic transition frequency modulation
through the atom-cavity coupling. However, the photon generation process is
not steady because after several photons have been created the third and
further terms in (\ref{PDC}) become important, and the photon generation is
interrupted. Nevertheless, the Hamiltonian (\ref{PDC}) shows that it is
possible to simulate DCE and generate several photons from vacuum using a
single artificial atom. This is illustrated in Fig. 2, where we show the
exact dynamics of $\left\langle n\right\rangle $ and $P_{e}$ \textit{vs}.
time for the DCE resonance, $f_{t}=\sin \eta _{DCE}t$ with $\xi =\delta $,
for the initial state $|g,0\rangle $ and the parameters $\Omega _{0}/\omega
=1.4$, $g_{0}/\omega =2\cdot 10^{-2}$, $\varepsilon /\omega =0.4$. We also
show the curve $\left\langle n\right\rangle _{DCE}=\sinh ^{2}\left(
2\left\vert \delta \theta \right\vert t\right) $, which gives the expected
mean photon number for the DCE Hamiltonian (\ref{DCE}), demonstrating that
for initial times the exact dynamics can be described by the DCE
Hamiltonian. However, after the creation of a few photons the atom acquires
a finite probability of being excited, and the dynamics starts to deviate
from the DCE Hamiltonian. The photon generation is interrupted and restarts
again as time goes on, respecting the limit $g_{0}\sqrt{\left\langle
n\right\rangle }/\left\vert \Delta _{-}\right\vert \ll 1$, and the behavior
of $P_{e}$ resembles the one of $\left\langle n\right\rangle $.
\begin{figure}[tbp]
\begin{center}
\includegraphics[width=.49\textwidth]{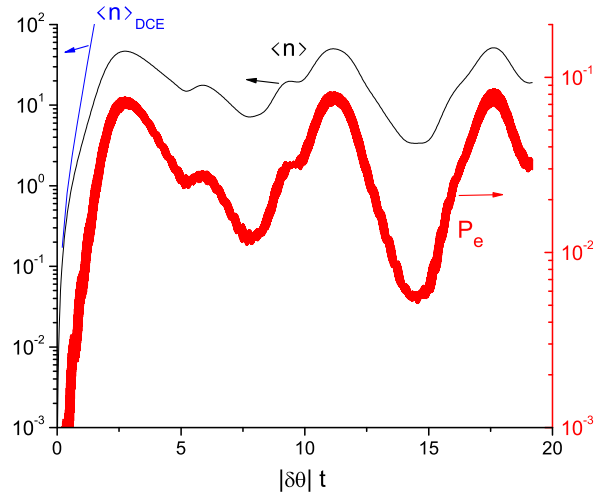} {}
\end{center}
\caption{Exact dynamics of $\left\langle n\right\rangle $ and $P_{e}$
\textit{vs}. time in the dispersive regime for the DCE resonance\textbf{. }$%
\left\langle n\right\rangle _{DCE}$ is the mean photon number for the DCE
Hamiltonian (\protect\ref{DCE}).}
\end{figure}

This phenomenon may be qualitatively understood as follows. In the
dispersive regime the atom acts as an effective non-linear capacitance \cite%
{PRA69-062320}, pulling the cavity frequency to
\begin{equation*}
\tilde{\omega}\left( t\right) \approx \omega +\frac{\sigma _{z}g_{0}^{2}}{%
\Delta _{-}+\varepsilon f_{t}}\approx \left( \omega +\sigma _{z}\delta
\right) -\sigma _{z}\delta \frac{\varepsilon }{\Delta _{-}}f_{t}\,.
\end{equation*}%
Consequently, one expects that the periodic modulation of $f_{t}$ with the
modulation frequency close to $\eta \approx 2\left( \omega \pm \delta
\right) $ would lead to DCE \cite{book,pp}, for which the photons are
generated as long as the modulation is present. The energy $2\omega $
necessary to create pairs of photons is provided through the atomic
frequency modulation and the resonance shift $\xi $ must be adjusted \cite%
{foot} in order to get a constructive interference on the cavity field \cite%
{book,JRLR26-445}. However, as time goes on the atom gets entangled with the
field, acquiring a finite probability of being excited through photon
absorption, and the photon generation cannot continue steadily due to the
loss of constructive interference. This is different from the usual DCE
situation, where the properties of the macroscopic linear, lossless and
nondispersive dielectric medium inside the cavity are modulated \cite{Law,Law1,jj,jj2},
and the field does not get entangled with individual atoms.

\section{Discussion and conclusions}

In the previous sections we have considered only the first order resonances.
In general, the $K$-th order resonances occur for an integer $K$ when%
\begin{equation*}
\eta =\eta _{i}^{\left( K\right) }\equiv K^{-1}\eta _{i},
\end{equation*}%
where $\eta _{i}$ stands for the AJC, JC and DCE resonances. In this case
one recovers the previous result upon substitutions $\delta \rightarrow
\delta _{K}\approx \delta +g_{0}^{2}/\Delta _{+}\cdots $ and $\theta
\rightarrow \theta _{K}$, as follows from the expressions (\ref{a1}) and (%
\ref{eee}). Now the effective dispersive shift $\delta _{K}$ contains the
contribution of many terms, among them the Bloch-Ziegert \cite{Klimov} shift
$g_{0}^{2}/\Delta _{+}$ and powers of $\left( \varepsilon /\eta \right) $. $%
\theta _{K}$ contains contributions due to the non-harmonic shape of the
pulse, $\Lambda _{k>1}$, as well as due to the powers of $\left( \varepsilon
/\eta \right) $%
\begin{equation*}
\theta _{K}=\Lambda _{K}\,\varepsilon /\eta +\cdots +\left( \Lambda
_{1}\,\varepsilon /\eta \right) ^{K}/K!+\cdots .
\end{equation*}%
However, to employ the higher order resonances the effective dispersive
shift $\delta _{K}$ should be carefully evaluated, otherwise there is a risk
of missing the exact modulation resonance, since $\left\vert \theta
_{K}\right\vert $ becomes smaller and there is less freedom in committing
small errors in the resonance shift $\xi $.

Our results may be easily translated to the situation where $\Omega \left(
t\right) =\Omega _{0}$ is constant and $g_{0}\left( t\right)
=g_{0}+\varepsilon f_{t}$ has a periodic time-dependence. In this case the
interaction Hamiltonian is%
\begin{equation}
H_{I}=g_{0}\left( t\right) \left( e^{i\left( \Omega _{0}-\omega \right)
t}a\sigma _{+}+e^{i\left( \Omega _{0}+\omega \right) t}a^{\dagger }\sigma
_{+}+h.c.\right) .  \label{boris}
\end{equation}%
If one expands $\exp \left( i\Xi _{\pm }\right) $ in Eq. (\ref{eee}) to the
first order in $\varepsilon /\eta $, the Hamiltonian (\ref{a1}) becomes
equivalent to the Hamiltonian (\ref{boris}), so the results obtained above
for $\Omega \left( t\right) $ also hold for $g_{0}\left( t\right) $ after
making appropriate substitutions. The main difference is that in the $\Omega
\left( t\right) $ case the higher order resonances occur due to the powers
of $\varepsilon /\eta $ and non-zero coefficients $\Lambda _{K}$, while in
the $g_{0}\left( t\right) $ case they are due to non-zero coefficients $%
\Lambda _{K}$ only. Finally, if the cavity frequency is modulated
periodically, with constant $\Omega $ and $g_{0}$, the AJC and JC resonances
also occur, besides the well known DCE resonance \cite{pp}.

The experimental verification of this scheme seems possible in circuit QED
architecture with superconducting qubits and coplanar waveguide resonators
\cite{PRA69-062320}, where one may adjust the system parameters \emph{in situ%
} via electric and magnetic fields, as demonstrated in \cite%
{N431-162,xx6,N449-443}. Moreover, several schemes to read out the cavity
and the atomic states are currently available \cite{N445-515,PRA69-062320,r2}
and under investigation \cite{PRA75-032329}. The main issue would be to
modulate periodically the atomic transition frequency with a stable
modulation frequency $\eta \sim 10$\thinspace GHz, what is within
experimental reach \cite{xx6}. One could also use this scheme to couple $M$
identical qubits (e.g. superconducting 2-level atoms \cite{xx6,N449-443} or
a cloud of polar molecules \cite{NP2-636}) to the same cavity mode and
modulate the frequency of $M$ atoms simultaneously, since in this case the
effective coupling increases to $\sqrt{M}g_{0}$.

One important point we did not analyze here is the dissipation and
decoherence of both the artificial atom and the cavity due to the noisy
solid state environment \cite{PRA75-032329,lula}. Recent experiments
achieved experimental values $\left\{ \kappa /\omega <10^{-4},\gamma /\omega
<10^{-3},\gamma _{ph}/\omega <10^{-3}\right\} $ \cite{N445-515}, where $%
\kappa $ is the cavity decay rate, $\gamma $ is the atomic decay rate and $%
\gamma _{ph}$ is the atomic pure dephasing rate. To deal with dissipation in
a qualitative manner, we compare the rates of the photon production from
vacuum for each resonance to the dissipation rates. We take\ the current
experimental value of the coupling constant $g_{0}/\omega \approx 2\cdot
10^{-2}$ \cite{N445-515} and assume $\varepsilon /\omega \sim \Delta
_{-}/\omega \sim 10^{-1}$ to make the estimative. The photon creation rate
for the first order DCE resonance is roughly $\left\vert \delta \theta
\right\vert /\omega \sim 10^{-4}$, and for the first order AJC resonance $%
\left\vert g\theta \right\vert /\omega \sim 10^{-3}$. Both these values are
larger or of the order of magnitude of the dissipation/decoherence rates, so
the photon production due to modulation of $\Omega \left( t\right) $ seems
possible in the future.

In conclusion, we analyzed the nonstationary circuit QED system in which the
atomic transition frequency has a small periodic modulation in time,
prescribed externally. In the dispersive regime, under the modulation
`resonances' the dynamics can be effectively described by the
Anti-Jaynes-Cummings, Jaynes-Cummings or the dynamical Casimir effect
Hamiltonians. Moreover, in the resonant atom-cavity regime, under the
corresponding `resonance', an entangled state with two photons can be
created from the vacuum state $|g,0\rangle $, analogously to the dynamical
Casimir effect in a vibrating cavity containing a two-level atom. This study
illustrates the importance of the anti-rotating term in the Rabi
Hamiltonian, neglected in the Jaynes-Cummings model -- here this term is
responsible for photon generation from vacuum and field amplification. As
applications, this scheme can be used to verify photon creation from vacuum
in nonstationary circuit QED due to an analog of the dynamical Casimir
effect using a single atom, as well as off-resonant transitions between the
states $\left\{ |g,m\rangle ,|e,m\pm 1\rangle \right\} $ and generation of
entangled states with several photons.

\begin{acknowledgments}
Work supported by FAPESP (SP, Brazil) Grant No. 04/13705-3, and by ITAMP at
Harvard University and Smithsonian Astrophysical Observatory. This work was
(partially) supported by National Science Foundation.
\end{acknowledgments}


\begin{thebibliography}{99}
\bibitem{foot} One could change $\xi $ adiabatically in time to enhance the
resonance.
\end{thebibliography}
\end{document}